\documentclass[showpacs,preprintnumbers,amsmath,amssymb,aps,twocolumn,superscriptaddress,bibnotes]{revtex4}

\usepackage{amsfonts}
\usepackage[dvips]{graphicx}
\usepackage{type1cm}
\usepackage{color}
\usepackage{bm}

\newcommand{\eqnref}[1]{Eqn.~\ref{#1}}

\newcommand{\figref}[1]{Fig.~\ref{#1}}
\newcommand{\diffd}{\text{d}}
\newcommand{\e}[1]{\text{e}^{#1}}

\renewcommand{\vec}[1]{\mathbf{#1}}
\newcommand{\punc}[1]{\,#1}

\newcommand{\neweqnline}{\nonumber\\}

\newcommand{\unit}[1]{\text{#1}}

\begin{document}
\title{A repulsive atomic gas in a harmonic trap on the border of
  itinerant ferromagnetism}
\author{G.J.~Conduit}
\email{gjc29@cam.ac.uk}
\affiliation{Cavendish Laboratory, 19, J.J.~Thomson Avenue, Cambridge, CB3 0HE. UK}
\author{B.D.~Simons}
\affiliation{Cavendish Laboratory, 19, J.J.~Thomson Avenue, Cambridge, CB3 0HE. UK}

\date{\today}

\begin{abstract}
Alongside superfluidity, itinerant (Stoner) ferromagnetism remains one
of the most well-characterized phases of correlated Fermi systems. A 
recent experiment has reported the first evidence for novel phase 
behavior on the repulsive side of the Feshbach resonance in a two-component 
ultracold Fermi gas. By adapting recent theoretical studies to the atomic 
trap geometry, we show that an adiabatic ferromagnetic transition would take 
place at a weaker interaction strength than is observed in experiment. 
This discrepancy motivates a simple non-equilibrium theory that takes account 
of the dynamics of 
magnetic defects and three-body losses. The formalism developed displays 
good quantitative agreement with experiment.
\end{abstract}


\maketitle

The transition between ferromagnetism and paramagnetism with increasing 
temperature is considered to be a canonical example of a continuous phase 
transition, and the phenomenon has been firmly established in many materials. 
The ability to tune the critical temperature through varying pressure 
presents experimentalists with an opportunity to investigate the novel 
many-body physics predicted to arise in the vicinity of a quantum critical 
point~\cite{Hertz}. However, whether this phase behavior derives from soft 
magnetic fluctuations, which are believed to drive the Stoner transition first
order~\cite{Abrikosov58,Duine05,Conduit08}, or is a consequence of coupling 
to auxiliary degrees of freedom such as lattice vibrations remains a subject 
of considerable interest and debate~\cite{QuantumConundrum}. The control 
afforded by Feshbach resonance phenomena in ultracold degenerate Fermi gases
presents a controlled 
platform from which to explore strongly correlated repulsive Fermi systems, 
including collective effects from non-equilibrium dynamics and spin
textures~\cite{Coleman08,LeBlanc09}, and quantum critical 
phenomena~\cite{Hertz}.

Although a few early experiments~\cite{Gupta03,Bourdel03} conducted on the 
repulsive side of the resonance hinted at ferromagnetic 
behavior~\cite{Duine05}, these investigations were hindered by the
challenges posed by the cold atomic gas set-up, with the fixed relative 
populations of particles, trap confinement, atom loss through three-body 
interactions, and non-equilibrium physics, rendering the conclusive 
identification of ferromagnetism impossible. However, in a recent study, 
Jo~\emph{et~al.}~\cite{Jo09} succeeded in observing the first strong 
evidence for novel phase behavior, consistent with itinerant ferromagnetism 
in an atomic gas of $\sim6.5\times10^{5}$ $^{6}\text{Li}$ atoms. To overcome 
the obstacle of atom loss through three-body interactions the experiment 
was carried out under non-adiabatic conditions, with the atoms 
prepared in the disordered non-ferromagnetic state and the magnetic field ramped
to the repulsive side of the resonance in $4.5\unit{ms}$ and then held fixed 
for a further $2\unit{ms}$. To assess the viability of the experimental
design, it is crucial to have detailed predictions for the expected 
phase behavior in the trap geometry. In the following, we will adopt
two stands of investigation. Firstly, it is important to understand
what we expect to see in an idealized equilibrium trap geometry. This 
study provides a benchmark both to assess the current experiment and 
to guide future studies. However, while this analysis achieves a 
qualitative agreement with experiment, it also
exposes important discrepancies highlighting the need to consider
non-equilibrium effects. In the second part of the letter
we turn to address non-equilibrium dynamics to critically analyze the 
experimental observations and establish strong quantitative agreement between
theory and experiment.

\emph{An atomic gas at equilibrium} provides an ideal platform from which 
to analyze experiment. For a two-component Fermi gas with a local contact 
interaction and equal masses, the Stoner Hamiltonian is defined by 
$H= \int\text{d}^3x\,[\sum_\sigma a_\sigma^\dagger (\hat{\vec{p}}^2/2m) a_\sigma 
+g a_\uparrow^\dagger a_\downarrow^\dagger a_\downarrow a_\uparrow]$, where the atoms 
are distinguished by a pseudospin index $\sigma\in\{\uparrow, \downarrow\}$. 
Taking into account all contributions to second order in $g$, the 
energy density of the Stoner Hamiltonian for a spatially uniform system 
can be expressed as~\cite{Abrikosov58,Duine05,Conduit08},
\begin{eqnarray}
&\varepsilon&=\frac{1}{V}\sum_{\vec{k},\sigma}\epsilon_{\vec{k}}
n_{\sigma}^{\text{F}}(\epsilon_{\vec{k}})+\frac{2k_{\text{F}}a}
{\pi\nu V^{2}}N_{\uparrow}N_{\downarrow}-\frac{2}{V^{3}}\left(
\frac{2k_{\text{F}}a}{\pi\nu}\right)^{2}\nonumber\\
&\times&\!\!\!\!\sum_{\vec{k}_{1,2,3,4}}\!\frac{n_{\uparrow}^{\text{F}}
(\epsilon_{\vec{k}_{1}})n_{\downarrow}^{\text{F}}(\epsilon_{\vec{k}_{2}})
\left[n_{\uparrow}^{\text{F}}(\epsilon_{\vec{k}_{3}})+n_{\downarrow}^{\text{F}}
(\epsilon_{\vec{k}_{3}})\right]}{\epsilon_{\vec{k}_{1}}+\epsilon_{\vec{k}_{2}}
-\epsilon_{\vec{k}_{3}}-\epsilon_{\vec{k}_{4}}}\punc{,}
\label{eqn:TheAFMFreeEnergy}
\end{eqnarray}
where
$n_{\sigma}^{\text{F}}(\epsilon)=1/[1+\e{\beta(\epsilon-\mu_{\sigma})}]$
denotes the Fermi distribution, with spectrum $\epsilon_{\vec{k}}$,
chemical potential $\mu_{\sigma}$, and reduced temperature $\beta$,
$N_{\sigma}$ is the number of particles, $\nu$ is the density of
states at the Fermi surface, $k_{\text{F}}$ is the Fermi wave vector
of the corresponding non-interacting system, and the scattering length
$a$ fully characterizes the strength of the contact interaction close
to resonance. We also set $\hbar=m=1$. Retaining only the leading
interaction correction, $\mathcal{O}(k_\text{F}a)$,
\eqnref{eqn:TheAFMFreeEnergy} recovers the conventional mean-field
Stoner theory, which predicts that the transition remains second order
down to zero temperature. However, the soft transverse magnetic
fluctuations~\cite{Belitz98} encoded in the second order
term~\cite{ConduitGreenSimons09} have the effect of driving the
ferromagnetic transition first
order~\cite{Abrikosov58,Duine05,Conduit08} at low temperature.

To address the non-uniform atomic trap geometry, we must determine the
free energy density,
$f=\varepsilon-\mu[n_{\uparrow}(\vec{r})+n_{\downarrow}(\vec{r})]$. Here
 the Lagrange multiplier $\mu$ enforces the constraint of a fixed
total particle number imposed by the trap geometry. It is also
convenient to rotate the basis to the axis of net magnetization
$s\in\{+,-\}$. To analyze the spatially inhomogeneous atom
distribution we invoke a local density approximation which allows the
variational minimization $\frac{\delta f}{\delta n_{s}(\vec{r})}$. 
This leads to two equations for the effective local chemical potentials
$\mu_{\pm}(\vec{r})$ of the majority and minority species in the
rotated basis that must be solved self-consistently,
\begin{eqnarray}
 &&\!\!\!\!\!\!\!\mu_{\pm}(\vec{r})\!=\!\mu\!-\!V(\vec{r})\!-\!\frac{2k_{\text{F}}a}{\pi\nu}n_{\mp}^{\text{F}}(\vec{r})\!+\!2\!\left[\!\frac{2k_{\text{F}}a}{\pi\nu}\!\right]^{2}\!\!\!\!\sum_{\vec{k}_{1,2,3,4}}\!\!\!\!n_{\mp}^{\text{F}}(\epsilon_{\vec{k}_{2}})\times\neweqnline
&&\!\!\!\!\!\!\!\frac{n_{\pm}^{\text{F}}(\epsilon_{\vec{k}_{1}})\delta(\epsilon_{\vec{k}_{3}}\!\!-\!\mu_{\pm})\!+\![n_{\pm}^{\text{F}}(\epsilon_{\vec{k}_{3}})\!+\!n_{\mp}^{\text{F}}(\epsilon_{\vec{k}_{4}})]\delta(\epsilon_{\vec{k}_{1}}\!\!-\!\mu_{\pm})}{\epsilon_{\vec{k}_{1}}+\epsilon_{\vec{k}_{2}}-\epsilon_{\vec{k}_{3}}-\epsilon_{\vec{k}_{4}}}\punc{,}
 \label{eqn:EulerLagrangeEqn}
\end{eqnarray}
within the external confining potential
$V(\vec{r})=\omega \vec{r}^{2}/2$. To provide a reference with
experiment~\cite{Jo09}, we describe the interactions in terms of a
dimensionless interaction strength expressed in terms of the Fermi
wave vector $k_{\text{F}}^{0}a$ at the center of a 
trapped non-interacting Fermi gas.

\begin{figure}
 \centerline{\resizebox{0.85\linewidth}{!}{\includegraphics{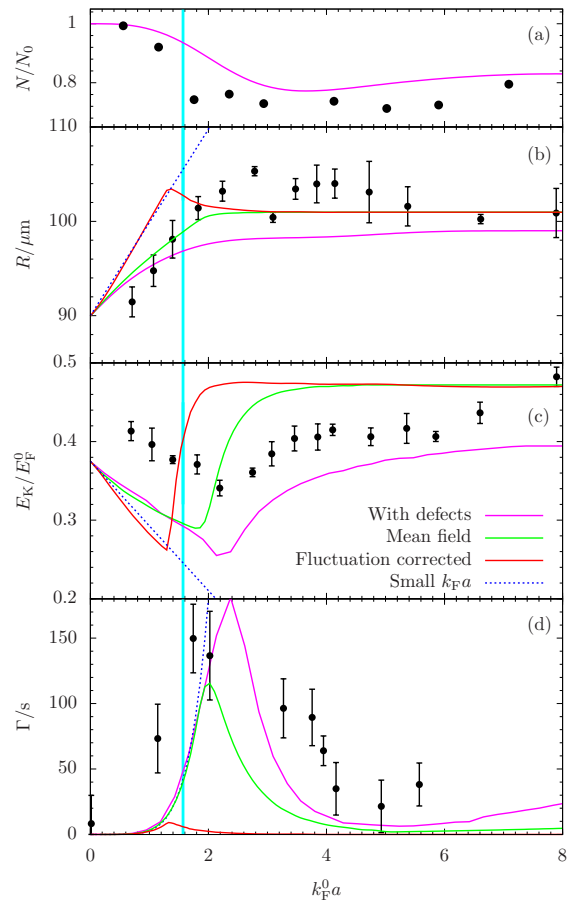}}}
 \caption{(Color online) (a) Atoms remaining after the hold time, (b)
   RMS cloud size, (c) release energy, and (d) loss rate with the
   dimensionless interaction parameter $k_{\text{F}}^{0}a$ shown on
   the primary x-axis in the mean-field case (green line), with
   fluctuation corrections (red line), and with defects (magenta
   line). The dashed blue line shows the trend in the small
   $k_{\text{F}}^{0}a$ limit, the cyan line is the Stoner criterion,
   and the experimental points of Ref.~\cite{Jo09} are also highlighted.}
 \label{fig:PlotCombined}
\end{figure}

We first address the behavior of the cloud size across the range of
interaction strengths. \figref{fig:PlotCombined}(b) confirms that, for
weak interactions, the RMS cloud size follows the 
universal scaling relationship $R_{0}^{\text{RMS}}(1+\frac{1024}{945\pi^{2}}
k_{\text{F}}^{0}a)$ that can be derived from
\eqnref{eqn:EulerLagrangeEqn}. However, fluctuation corrections soon increase
the cloud pressure causing it to inflate rapidly to its fully
polarized value $2^{1/6}R_{0}^{\text{RMS}}$. As cloud density is
suppressed by pressure, the onset of ferromagnetism at the center of
the trap takes place at an enhanced interaction strength compared with 
the uniform system. Secondly, in \figref{fig:PlotCombined}(c)
we study the total kinetic energy of the atoms, which is
in agreement with the mean-field prediction of LeBlanc \emph{et
al}.~\cite{LeBlanc09}. With increasing interaction strength the
local density falls so that the kinetic energy decreases to
$\frac{3}{8}E_{\text{F}}^{0}(1-\frac{2048}{945\pi^{2}}k_{\text{F}}^{0}a)$.
The fluctuation corrections increase the cloud pressure and reduce the
density so driving the kinetic energy downwards still further.
The release energy
recorded by experiment extrapolated to zero interaction strength is
$\sim0.45E_{\text{F}}^{0}$, which is marginally higher than the
analytical prediction of $0.375E_{\text{F}}^{0}$.                    
This discrepancy could arise through the recovery of
interaction energy upon release from the trap, remnant eddy currents
in the coils, or non-equilibrium effects.
The final experimental probe that we address here is the atom loss
rate due to three-body recombination. Integrating the loss rate
$\Gamma=\Gamma_{0}(k_{\text{F}}a)^{6}\int n_{+}(\vec{r})
n_{-}(\vec{r})[n_{+}(\vec{r})+n_{-}(\vec{r})]\diffd^{3}r$~\cite{Petrov03}
over the trap yields the variation with interaction strength shown in
\figref{fig:PlotCombined}(d). The mean-field case agrees with the
prediction of LeBlanc {\it et al}.~\cite{LeBlanc09}. The initial rapid
rise in the loss rate can be attributed to the increase in the
$(k_{\text{F}}^{0}a)^{6}$ coefficient, through the universal relation
$\Gamma/\Gamma_{0}\approx7(k_{\text{F}}^{0}a)^{6}\mu^{6}/288
\pi^{4}\omega^{3/2}$. The later decay in the loss rate is due to the
suppression of the product $n_{+}(\vec{r})n_{-}(\vec{r})$ with the
onset of magnetization.

\begin{figure}
 \centerline{\resizebox{0.85\linewidth}{!}{\includegraphics{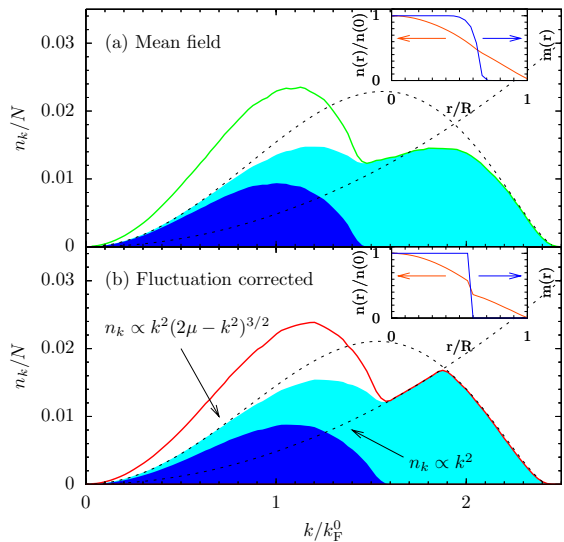}}}
 \caption{(Color online) The momentum distribution $n_{k}/N$ of a half
   net polarized atomic gas in (a) the mean-field case and (b) with
   fluctuation corrections. The cyan filled curve shows the population
   of majority spin particles and the blue filled curve shows the
   minority spin particles. The dotted curves show the expected
   distribution for the majority spin particles in a harmonic trap,
   $n_{k}\propto k^{2}(2\mu-k^{2})^{3/2}$, and simply due to the
   density of states, $n_{k}\propto k^{2}$. The inset curves show the
   radial distribution of atoms (orange, primary y-axis) and
   magnetization (blue, secondary y-axis).}
 \label{fig:MomDistib}
\end{figure}

Although there is reasonable qualitative agreement between theory and 
experiment, a marked divergence arises in
the experimental prediction of the interaction strength at the onset of 
ferromagnetism at $k_{\text{F}}a\approx2.2$. The theoretical prediction 
from mean-field theory is $k_{\text{F}}a\approx1.9$, whereas it is at 
$k_{\text{F}}a\approx1.1$ if fluctuation 
corrections are taken into account. This discrepancy 
prevents us from drawing a definitive conclusion on whether the 
transition is first or second order. However, in future experimental 
studies, it should be possible to gain insight into this question  
by studying the momentum distribution of the atoms shown in
\figref{fig:MomDistib}. The latter can be directly measured by
studying the spatial distribution following a ballistic
expansion~\cite{KetterleReview08}. A single non-interacting atomic
species has the momentum distribution $n_{k}\propto
k^{2}(2\mu-k^{2})^{3/2}$. 
In the partially polarized regime the atomic gas will be ferromagnetic
in the high density regions -- at the center of the trap. Since the
transition from an unpolarized to a fully polarized gas takes place over a 
small range of interaction strengths, the atomic gas will be
partially polarized over a small range of radii. The remainder of the
gas is unpolarized.
The thin shell of partially polarized gas gives rise to a sharp
double peak structure in the momentum distribution 
in both the mean-field case and also when
fluctuation corrections are taken into account. However, the
mean-field ferromagnetic transition takes place over a larger range of
radii than the fluctuation corrected case, which smears out the double
peak feature.

To assess whether the observed experimental phenomenology \cite{Jo09} is
consistent with the development of itinerant ferromagnetism, it is
important to understand the source of the discrepancy in the predicted 
interaction strength. In doing so, we will exploit the marginally
adiabatic nature of the experiment to explore the quench dynamics of the 
transition. This necessitates taking into account two contributing 
factors: how the ferromagnetic state condenses out of the normal phase, 
and the renormalization of the interaction strength due to three-body 
losses. The emergence of the ferromagnetic state is non-trivial since the 
propagation of the condensed spin alignment is bounded by the spin wave 
velocity. The
quench propels the system deep into the ferromagnetic regime, so
monopole defects are condensed out from the paramagnetic 
phase~\cite{Langer92,BrayReview}. As the defects grow they mutually 
annihilate, delaying the formation of the ferromagnetic phase, which 
in turn increases the effective interaction strength required to 
observe ferromagnetic phenomena.
The second component of our dynamical study are three-body 
interactions. These reduce the atom density and therefore
abate the effective interaction strength $k_{\text{F}}a$, 
which to compensate must be artificially raised to observe 
ferromagnetism. The interplay between these two dynamical effects 
is expected to raise the interaction strength required to observe 
ferromagnetism to be more in accord with the experimental findings.
We now detail how to separately incorporate these effects
into a description of the atomic gas, before merging them into a
single formalism. Since
the characteristic timescale associated with temperature,
$0.5\unit{ms}$, is similar to time scales of the Ketterle experiment
\cite{Jo09}, we can treat the atomic gas as if it were at 
zero temperature. Due to the short time-scale of
the quench, for simplicity we can focus attention on just the 
mean-field component of the theory~\cite{Demler09} and make the 
approximation that the ferromagnetic transition takes place at the 
Stoner criterion $k_{\text{F}}a_{\text{CRIT}}=\pi/2$. 

\emph{The ferromagnetic quench} deep beyond the spinodal line
leads to the condensation of topological monopole
defects~\cite{BrayReview}. In each defect the 
spins radiate out from a central core with magnetization 
$\phi(r>\xi)=\phi_{0}(1-\xi^{2}/r^{2})$, where $\xi=1/k_{\text{F}}
\sqrt{k_{\text{F}}a/k_{\text{F}}a_{\text{CRIT}}-1}$ is the healing length.
A uniform atomic gas would be only partially polarized over a range of 
interaction strengths $\Delta\approx0.29$, 
so therefore during the field ramp over time 
$t_{\text{ramp}}$ to a target interaction
strength $k_{\text{F}}a_{\text{AIM}}$, the gas will be partially 
polarized for the time $t_{\text{ramp}}\Delta/k_{\text{F}}a_{\text{AIM}}$. 
The study of Babadi~\cite{Demler09} shows that after this characteristic
quench time the defects will 
adopt an initial size $L_{0}=[(\hbar/mk_{F}^{0\,\,{3/2}}/
6\pi^{2})t_{\text{ramp}}\Delta/k_{\text{F}}a_{\text{AIM}}]^{1/3}$. 
Subsequently, the defects grow and annihilate as they compete to select 
the equilibrium spin. This expansion of the defects can be 
summarized by the growth law of $\sqrt{\hbar t/m}$ \cite{Langer92}.
Combined with the initial condensed size of the defects, we 
therefore model the defect size by $L(t)=\lambda(L_0+\sqrt{\hbar t/m})$, 
with $\lambda$ being an unknown dimensionless constant that we include 
to account for the system dependent dynamics that go beyond the current
analysis.

Having captured defect
growth in the model, we now incorporate it into a 
formalism that also accounts for the atom loss due to three-body interactions. 
We employ the standard atom loss rate formula~\cite{Petrov03} to describe 
three body recombination $\dot{n}=-111(na^{3})^{2}\bar{\epsilon}n_{\uparrow}
n_{\downarrow}/\hbar n$,
where $\bar{\epsilon}=4.56\hbar^{2}n^{2/3}/m$ is the 
average kinetic energy. The fall in loss rate due to Pauli blocking is 
described 
by a geometric term that can be expressed in terms of the magnetization 
$\phi$ through
$n_{\uparrow}n_{\downarrow}/n^{2}=(n^{2}-\phi^{2})/4n^{2}$. As the spins 
within a defect are not parallel, three-body recombination occurs within 
a defect, which can be expressed 
through the geometric term as 
$n_{\uparrow}n_{\downarrow}/n^{2}=\xi^{2}/2r^{2}+\mathcal{O}(\xi^{4})$. 
Integrating this loss over one defect predicts a net loss rate of
\begin{eqnarray}
&&\frac{4\pi L^{3}(t)}{3}\frac{\diffd n}{\diffd t}\!=\!-111n(na^{3})^{2}
\frac{\bar{\epsilon}}{\hbar}\int_{L(0)}^{L(t)}\!\!\frac{\xi^{2}}{4r^{2}}
4\pi r^{2}\diffd r\neweqnline
&&\frac{\diffd n}{\diffd t}\!=\!-0.144\frac{\hbar}{m}n^{5/3}
(k_{\text{F}}a)^{6}\![L(t)\!-\!L(0)]\xi^{2}L^{-3}(t)\punc{.}
\end{eqnarray}
Starting in the disordered regime, following the magnetic field ramp
we numerically
propagate the atom loss rate forwards in time. If the system breaks
into the ferromagnetic regime, $L(t)$ starts to grow as the defects
expand and annihilate. Should atom loss cause the system to reenter
the normal regime we immediately revert to the $L=0$ limit.
To determine the atomic distribution within the trap we
self-consistently evaluate the local effective potential following
the method employed for the uniform case in \eqnref{eqn:EulerLagrangeEqn}.

In \figref{fig:PlotCombined}(a) we first analyze the population of remnant 
atoms following the field ramp. At weak interaction strengths the number of 
atoms lost to three body recombination scales as $\sim(k^{0}_{\text{F}}a)^{6}$. 
The onset of ferromagnetism reduces the atom loss, as the more rapid onset of
ferromagnetism and smaller core length conspire to cut losses. Deep within
the ferromagnetic regime, around $80\%$ of the atoms remained in the 
experiment~\cite{Jo09}, which was used to calibrate the constant 
$\lambda=5$. Having determined the
dimensionless constant, we now monitor the ramifications of this choice
on the growth of the defects. The ratio of the defect size to the core
radius, $L/\xi$, after the hold time, increases with $k_{\text{F}}a$
due to the proliferation of defects commencing earlier and the falling
core radius. The rise of $L$ leads to a fall in the number of defects
remaining after the field hold to $\sim8$. 
However, no firm evidence has
yet been reported for domain formation in experiment.

Now that we have calibrated the theory against the experimental
results we can now proceed to compare the other physical observables
against the experimental data. We first look at the RMS cloud
size in \figref{fig:PlotCombined}(b).
In the weakly repulsive regime the cloud size increases with
interaction strength due to the enhanced pressure between
atoms. However, at interaction strengths $1<k_{\text{F}}^{0}a<2.2$ the
increasing atom loss dominates, causing the cloud to shrink. The
characteristic minimum formed is in accord with the
experimental results, and critically is delayed until the
interaction strength $k_{\text{F}}^{0}a\approx2.2$
both because the loss of atoms hinders
the onset of ferromagnetism, and since on the border of ferromagnetism
the longer core length increases atom loss. Upon ferromagnetic
ordering the atom loss falls
off causing the cloud to re-expand, which eventually exceeds its
original radius.
With weak interactions the kinetic energy in \figref{fig:PlotCombined}(c)
falls both as atoms are lost
and the cloud dilates. Upon reaching the critical interaction strength
of $k_{\text{F}}^{0}a\approx2.2$ ferromagnetic ordering takes place and the kinetic energy rises as the
ordered atoms posses a larger Fermi surface. Across the whole range
of interaction strengths the theoretical prediction for the kinetic energy 
falls short of the experimental results, which could be due to the
gas being inherently out of equilibrium, recovery of some
interaction energy upon release from the trap, or remnant eddy currents
in the coils. At
high interaction strength the kinetic energy continues to rise as the
atom loss decreases, indicative of the upturn seen in the experimental
results.
Finally, in \figref{fig:PlotCombined}(d) we address the atom loss rate
which climbs strongly as $\sim(k^{0}_{\text{F}}a)^{6}$ before the
onset of ferromagnetism. Consistent with the other experimental
probes, the peak atom loss is delayed until
$k^{0}_{\text{F}}a\approx2.2$. Following the onset of
ferromagnetism the atom loss drops away more slowly than without
dynamic effects due to the necessity for the atom loss rate to grow
and the initial atom loss during the ramp time.

In conclusion we have performed a detailed critique of the results
of the first experimental signs of
ferromagnetism in a cold atom gas. Firstly, 
we demonstrated that the results are not consistent with the formation
of an equilibrium ferromagnetic phase. Secondly, we demonstrated how
the experiment indicates that the ferromagnetic phase is formed through the
condensation of defect-antidefect pairs that subsequently undergo 
mutual
annihilation, which delays the formation of the ferromagnetic phase
to an enhanced interaction strength
of $k_{\text{F}}a=2.2$.

We thank Eugene Demler, Jonathan Edge, Andrew Green, Zoran Hadzibabic,
and especially Gyu-Boong Jo and Wolfgang Ketterle for useful
discussions.


\begin{thebibliography}{99}

\bibitem{Hertz} J.A. Hertz. Phys. Rev. B {\bf 14}, 1165 (1976);
A.J. Millis. Phys. Rev. B {\bf 48}, 7183 (1993).

\bibitem{BelitzReview05}
D. Belitz, T.R. Kirkpatrick and T. Vojta. Rev. Mod. Phys. {\bf 77}, 579 (2005).

\bibitem{Abrikosov58}
A.A. Abrikosov and I.M. Khalatnikov. Soviet Phys. JETP {\bf 6}, 888 (1958);
F. Mohling. Phys. Rev. {\bf 122}, 1062 (1961).

\bibitem{Duine05}
R.A. Duine and A.H. MacDonald. Phys. Rev. Lett. {\bf 95}, 230403 (2005).

\bibitem{Conduit08}
G.J. Conduit and B.D. Simons. Phys. Rev. A {\bf 79}, 053606 (2009).

\bibitem{QuantumConundrum} R.B. Laughlin \emph{et al.}, Adv. Phys. {\bf 50}, 361 (2001).

\bibitem{Gupta03}
S.~Gupta \emph{et al.}, Science {\bf 300}, 1723 (2003).


\bibitem{Bourdel03} T.~Bourdel \emph{et al.}, Phys. Rev. Lett {\bf 91}, 020402 (2003).

\bibitem{Jo09}
G.-B.~Jo \emph{et al.}, Science {\bf 325}, 1521 (2009).

\bibitem{Coleman08}
 I.~Berdnikov, P.~Coleman and S.H.~Simon. Phys. Rev. B {\bf 79}, 224403 (2009)

\bibitem{LeBlanc09}
L.J.~LeBlanc \emph{et al.}, Phys. Rev. A {\bf 80}, 013607 (2009).

\bibitem{Belitz98}
D.~Belitz \emph{et al.}, Phys. Rev. B {\bf 58}, 14155 (1998).
M. Shimizu. Proc. Phys. Soc. {\bf 84}, 397 (1964);
D. Belitz, T.R. Kirkpatrick, and T. Vojta. Phys. Rev. B {\bf 55}, 9452 (1997);
J. Betouras, D. Efremov and A. Chubukov. Phys. Rev. B {\bf 72}, 115112 (2005);
D.V. Efremov, J.J. Betouras and A. Chubukov. Phys. Rev. B {\bf 77}, 220401(R) (2008).

\bibitem{ConduitGreenSimons09}
G.J. Conduit, A.G. Green and B.D. Simons. arXiv:0906.1347.

\bibitem{KetterleReview08}
W.~Ketterle and M.W.~Zwierlein. arXiv:0801.2500.

\bibitem{Petrov03}
D.S.~Petrov. Phys. Rev. A {\bf 67}, 010703(R) (2003).

\bibitem{JoPrivate}
G.-B.~Jo and W.~Ketterle (private communication).

\bibitem{Demler09}
M. Babadi, D. Pekker, R. Sensarma, A. Georges and E. Demler.
arXiv:0908.3483.

\bibitem{BrayReview}
A.J. Bray. Adv. Phys. {\bf 43}, 357 (1994).

\bibitem{Langer92}
J.S. Langer, {\it Solids far from equilibrium}, edited by C. Godr\`eche
(Cambridge University Press) (1992).

\end{thebibliography}
\end{document}